\documentclass[12pt]{article}

\usepackage{amsmath, amssymb, slashed}
\usepackage[dvips]{graphicx}
\usepackage{color}

\begin{document}

\begin{titlepage}
\begin{center}

\hfill IPMU12-0193\\

\vspace{1.5cm}
{\large\bf Why is the Supersymmetry Breaking Scale Unnaturally High?}
\vspace{1.0cm}

{\bf Brian Feldstein}$^{(a)}$,
and
{\bf Tsutomu T. Yanagida}$^{(a)}$

\vspace{1.0cm}
{\it
$^{(a)}${Kavli IPMU, University of Tokyo, Kashiwa, 277-8583, Japan}
}
\vspace{1.0cm}

\abstract{Evidence is mounting that natural supersymmetry at the weak scale is not realized in nature.   On the other hand, 
string theory suggests that supersymmetry may be present at some energy scale, and gauge coupling unification implies that that energy scale may be relatively low.  A puzzling question is then why nature would prefer a low, but not completely natural supersymmetry breaking scale.  Here we offer one possible explanation, which simultaneously addresses also the strong CP and $\mu$ problems.  We introduce an axion, and suppose that the Peccei-Quinn and supersymmetry breaking scales are connected.  If we further assume that R-parity is not conserved, then the axion is required to be dark matter, and  the Peccei-Quinn/supersymmetry breaking scale is required to be at least $\sim 10^{12}$ GeV.  Gravity mediation then yields scalar superpartners with masses of at least $\sim 100$ TeV.  The gauginos are likely to obtain loop-factor suppressed masses through anomaly mediation and higgsino threshold corrections, and thus may be accessible at the LHC.  The axion should be probed at phase II of the ADMX experiment, and signs of R-parity violation may be seen in the properties of the gauginos.  
}

\end{center}
\end{titlepage}
\setcounter{footnote}{0}

\section{Introduction}
The lack of any observation of superpartners or new flavor violating effects, as well as the observed Higgs mass of 125 GeV, suggest that if supersymmetry (SUSY) is realized in nature, the SUSY breaking scale is likely appreciably higher than the weak scale.  On the other hand, string theory suggests that supersymmetry may be realized at {\it some} scale in nature, and gauge coupling unification, naturalness and the presence of dark matter all point towards a low supersymmetry breaking scale.  In fact, if the scalar superpartner masses are perhaps 100-1000 TeV, with gauginos obtaining lighter 100-1000 GeV masses automatically by anomaly mediation \cite{anomaly}, then the resulting picture is remarkably consistent:  The heavy scalars appropriately raise the Higgs mass through loop corrections, suppress flavor changing neutral currents, and allow for the avoidance of collider constraints.  In such a scenario, supersymmetry breaking may be communicated by supergravity effects alone, without the need for additional structure.

In this picture- dubbed ``Pure Gravity Mediation" (PGM) \cite{Ibe1, Ibe2, LHCPGM}- gauge coupling unification remains successful, naturalness is significantly improved compared to a case with a higher SUSY scale (or no SUSY at all), and the lightest superpartner- the wino- may still serve as dark matter.  In spite of all of this, however, a crucial question nags:  If nature prefers to solve the hierarchy problem and therefore has a relatively low SUSY breaking scale, then why doesn't it go all the way?  There doesn't appear to be any fundamental physical constraint preventing the SUSY breaking scale from being lower, with a lighter dark matter particle and perfect naturalness.  In a landscape picture, it is difficult to imagine why a region with a low, but not completely natural SUSY scale would be a likely place for us to find ourselves. A small number of other open questions remain in this scenario as well, such as the reason for a $\mu$ parameter at least as small as the SUSY breaking masses and the reason for the suppression of strong CP violation, amongst others.

In this paper, we will show that the issues mentioned above- the $\mu$ problem, the strong CP problem, and the incomplete naturalness problem- may all be addressed simultaneously in a straightforward way.  Our key assumptions will be as follows:
\begin{itemize}
\item{The strong CP problem is solved by a Peccei-Quinn (PQ) symmetry \cite{PQ} under which the standard model quarks are charged, with an associated axion.}
\item{This same PQ symmetry forbids the presence of the $\mu$ term, and PQ symmetry breaking then leads to $\mu$ of an appropriate size.\footnote{See also for example \cite{PQmu}.}  Here we assume that the PQ charges are such that the $\mu$ parameter is suppressed by one power of the Planck scale.}
\item{The breaking of PQ symmetry triggers supersymmetry breaking, so that the two scales are generated to be roughly equal.}
\item{R-parity is not conserved.}
\end{itemize}

Our assumptions follow from a simple train of logic which begins with the hypothesis that an axion solves the strong CP problem.  In such a case,  PQ symmetry (assumed to be carried by the standard model quarks) automatically forbids the $\mu$ term, suggesting also a solution to the $\mu$ problem. 
 We are then faced with the prospect of generating two new scales in nature- the PQ scale, and the SUSY scale. 
 It is remarkable that the scale required for SUSY breaking in pure gravity mediation models $\Lambda_{SUSY} \sim \sqrt{ m_{3/2} M_{\rm pl}} \sim \sqrt{100 {\rm\ TeV} \times  10^{19} {\rm\ GeV}} \sim 10^{12}$ GeV happens to coincide with the PQ breaking scale typically needed for axionic cold dark matter \cite{PQDM}.  It is the simplest possibility then, that the scales are related, and moreover, that R-parity conservation is therefore not necessary in order to obtain a dark matter candidate.

  Assuming that the fundamental theory violates R-parity, the PQ scale, and therefore also the SUSY breaking scale {\it must} be at least of order $10^{12}$ GeV in order to obtain a dark matter candidate, as seems required for structure formation \cite{DMreq}.  We thus obtain a lower bound on the SUSY breaking scale of $\Lambda_{SUSY} \gtrsim 10^{12}$ GeV, implying gravity mediated gravitino and scalar masses $\gtrsim 100$ TeV.  While higher breaking scales are physically allowed,
they are disfavored by a resulting increase in fine-tuning- both in the weak scale, and also in the the initial displacement angle of the axion field to avoid dark matter over-production.   Note that if we took R-parity to be conserved, it would open up the possibility for LSP gaugino dark matter, with a lower SUSY breaking (and PQ) scale, and the mystery of an incompletely natural weak-scale would remain unsolved.\footnote{Indeed, there doesn't appear to be any problem with, for example, mixed bino-wino thermal relic dark matter with mass of perhaps $5-10$ GeV, and a few hundred GeV superpartner masses.}  We also note that our picture tends to result in a supersymmetry breaking spurion charged under PQ symmetry, which is thus unable to give masses to the gauginos directly, implying the anomaly mediated (plus higgsino threshold corrected) spectrum which we consider.

The outline of this paper is as follows:  In section \ref{bound}, we will briefly review some aspects of the pure gravity mediation scenario, before turning to discuss the details of our argument more thoroughly, as well as give an example of a simple model which can dynamically generate both the PQ and SUSY scales simultaneously.
In section \ref{pheno}, we will discuss the phenomenology of our scenario, including the requirements and constraints on the assumed R-parity violation.  Dark matter axions have a good chance to be observed in upcoming experiments, and the gauginos may be observable at the LHC, or a future linear collider.  The presence of R-parity violation may be suggested if the LSP is found to be a bino or gluino (since phenomenologically unacceptable contributions to the dark matter density would result without R-parity breaking).   Moreover, it may be possible to directly observe R-parity violating decays of the LSP, providing an explicit check of our framework, especially if accompanied by an observation of axionic dark matter.
We also point out that an interesting signal of our scenario would be to find a long lived wino LSP at a collider, but without a corresponding dark matter annihilation signal in cosmic rays.  This discrepancy could be particularly acute for relatively light LSP masses of perhaps a few hundred GeV.  We will summarize in section \ref{conclusions}.

\label{introduction}

\section{Bound on the SUSY Breaking Scale}

We will begin by reviewing the essential aspects of the pure gravity mediation scenario.  After supersymmetry breaking by an F term vev $F_{SUSY}$, the gravitino mass becomes
\begin{equation}
m_{3/2} = \frac{\left|F_{SUSY}\right|}{\sqrt{3} M_{\rm pl}},
\end{equation}
with $M_{\rm pl} = 2.4 \times 10^{18}$ GeV.  The scalar superpartners obtain masses at the same order.  The $\mu$ parameter, and hence the scale of the higgsino masses, in general depends on the details of the solution to the $\mu$ problem, but has been taken in
previous studies to be of order the gravitino mass as well.  This will indeed be the case in the specific scenario that we consider here.  The gauginos then obtain masses at 1-loop
order through anomaly mediation and also through higgsino threshold corrections.  After including renormalization group running effects as well, one obtains \cite{LHCPGM}
\begin{eqnarray}
\label{eq:gluino}
m_{\rm gluino} &\simeq&
2.5\times (1 - 0.13 \, \delta_{32} - 0.04 \, \delta_{\rm SUSY})
\times 10^{-2} \, m_{3/2},
\\ \label{eq:wino}
m_{\rm wino} &\simeq&
3.0\times(1 - 0.04 \, \delta_{32} + 0.02 \, \delta_{\rm SUSY})
\times 10^{-3} \, (m_{3/2} + L),
\\ \label{eq:bino}
m_{\rm bino} &\simeq&
9.6\times(1 + 0.01 \, \delta_{\rm SUSY})
\times 10^{-3} \, (m_{3/2} + L/11),
\end{eqnarray}
where $\delta_{\rm SUSY} = \log[M_{\rm SUSY}/100 \,{\rm TeV}]$, with $M_{SUSY}$ the scalar superpartner mass scale, here of order $m_{3/2}$. $\delta_{32}$ denotes $\delta_{32} = \log[m_{3/2}/100 \, {\rm TeV}]$ for the gluino, and $\delta_{32} = \log[(m_{3/2} + L) /100 \,{\rm TeV}]$ for the wino. The terms proportional to $m_{3/2}$ represent the anomaly mediated contributions, while those proportional to $L$ are the higgsino threshold contributions, with $L$ given by
\begin{eqnarray}
L \equiv \mu \sin 2\beta
\frac{m_A^2}{|\mu|^2 - m_A^2} \ln \frac{|\mu|^2}{m_A^2}
\label{eq:L}.
\end{eqnarray}
$m_A$ is the CP odd Higgs mass and $\tan{\beta}$ is the ratio of Higgs vevs as usual.  As discussed in reference \cite{Ibe1}, if $\mu$ is of the order of the gravitino mass along with the other soft mass parameters, then $L/m_{3/2}$ is generically of order 1.  The wino mass therefore obtains comparable contributions from both anomaly mediated effects and those of the higgsino threshold corrections.  Typically $L/{m_{3/2}}$ is required to be less than about 3 in order for the wino to be the LSP and provide a dark matter candidate, but since we will not assume gaugino dark matter in our discussions, this condition will not be necessary here.

We now assume the existence of a PQ symmetry under which  the Higgs fields and quark fields of the minimal supersymmetric standard model (MSSM) are charged.  We take the quark charges to be family universal for simplicity.  In this case, an $H_U H_D$ superpotential term is necessarily forbidden by PQ symmetry, arising only after
PQ symmetry breaking is communicated to the MSSM sector.  We will choose PQ charges such that the $\mu$ term is generated via a term in the superpotential
\begin{equation}
W \supset \frac{\kappa}{M_{\rm pl}} Q^2 H_U H_D,
\label{W}
\end{equation}
where $Q^2$ breaks PQ via an expectation value $\left<Q\right> = \Lambda_{PQ}/\sqrt{2}$.  We assume that concurrent with PQ symmetry breaking, a superfield $Z$ is triggered to obtain an F-term $F_Z$, breaking supersymmetry.   We define $\Lambda_{SUSY}^2 = F_Z =  \lambda  \Lambda_{PQ}^2$.  We take the SUSY/PQ breaking sector to not contain any small parameters which could lead to a significant hierarchy between $\Lambda_{SUSY}$  and $\Lambda_{PQ}$, so that $\lambda$ and $\kappa$ are of order 1.\footnote{Note that a fully strongly coupled theory would have $\lambda$ of order $4\pi$.}

  The strong anomaly  of the PQ symmetry leads to an interaction
\begin{equation}
\frac{\alpha_{QCD}}{8 \pi}(\theta_0 + \frac{a}{f_a}) G \tilde{G},
\end{equation}
where a is the axion field,  $\theta_0$ is the strong CP parameter,  G is the gluon field strength and
$f_a = \frac{ \Lambda_{PQ}}{3 N}$.  Here $N =\left| \frac{q_{H_U} + q_{H_D}}{q_0}\right|$, where $q_{H_U}$ and  $q_{H_D}$ are the Higgs PQ charges, and $q_0$ is the charge of the field which breaks PQ symmetry.  With the present example of 
the superpotential (\ref{W}), we have $N=2$, since $Q$ breaks $U(1)_{PQ}$.  On the other hand, for the dynamical breaking model to be discussed in section \ref{dynamical}, $Q^2$ will actually form a meson field $\sim \Lambda_{PQ} \Phi$, with $\Phi$ then obtaining a vev to break the symmetry.  In that case we will have $N=1$ instead.

QCD instantons generate a potential for $a$ of the form
\begin{equation}
\Lambda_{QCD}^4 \left(1- \cos{\left(\theta_0 + \frac{a}{f_a}\right)}\right),
\end{equation}
solving the strong CP problem.  This potential has a discrete symmetry under which $a\rightarrow a + 2 \pi f_a$, and since the full range for $a$ is actually $2 \pi \Lambda_{PQ}$ this is a $Z_6$ symmetry.
 This discrete symmetry results in
unacceptable cosmological domain walls \cite {walls} unless the reheating temperature $T_R$ is less than $\sim \Lambda_{PQ}$, and inflation leaves our Hubble volume with a fixed initial value for the axion $a = \Theta_i f_a$ \cite{wallsolve}.\footnote{Note that since PQ symmetry and supersymmetry are both broken in the same sector at the scale
 $\Lambda_{PQ}$, the saxion obtains a  mass of order $\Lambda_{PQ}$, and thus does not have
important effects on the cosmological history of the universe.}
  
Close to the time of the QCD phase transition, the axion field begins to oscillate around the minimum of its potential.  This leads to a contribution to the energy density of the universe today with \cite{PQDM}
\begin{equation}
\Omega_a h^2 \simeq .3 \left(\frac{f_a}{10^{12} {\rm GeV}}\right)^{7/6} \sin^2{\left(\Theta_i/2\right)}.
\end{equation}
If we assume that R-parity is broken, such that the dark matter must be composed of axions, then we obtain the requirement
\begin{equation}
\Lambda_{PQ} \simeq  \left(\frac{1}{\sin^{2}{\left(\Theta_i/2\right)}}\right)^{6/7} N\times10^{12}{\rm GeV},
\end{equation}
so that, with $\Theta_i < \pi$,
\begin{equation}
\Lambda_{SUSY} \gtrsim \sqrt{\lambda} N\times10^{12}{\rm GeV}.
\end{equation}
This gives a gravitino mass satisfying
\begin{equation}
m_{3/2} \gtrsim \lambda N^2 \times 350\ {\rm TeV}.
\end{equation}
\label{bound}

\subsection{Example Dynamical Symmetry Breaking Sector}
Here we will present one example of a symmetry breaking sector which could simultaneously  generate both the PQ and SUSY scales, and do so dynamically.
Our example will come from the work of \cite{dynamical}.
We will take $Q_i$ to be four chiral superfields in the fundamental representation of an $SU(2)_c$ gauge symmetry.  We will also take $Z_{ij}$ to be an anti-symmetric $4\times4$ matrix of $SU(2)_c$ singlets.  We assign PQ and R symmetry charges as in table \ref{table1}.
\begin{table}
\begin{center}
    \begin{tabular}{ | l | l | l |}
    \hline
     & PQ charge & R charge\\ \hline
    $Q_{1,2}$ & 1/2 & 0 \\ \hline
    $Q_{3,4}$ & -1/2 & 0 \\ \hline
    $Z_{12}$ & -1 & 2 \\ \hline
    other $Z_{ij}$ & 0 & 2 \\ \hline
    $Z_{34}$ & 1 & 2 \\ \hline
    $H_U$ & -1/2 & 1 \\ \hline
    $H_D$ & -1/2 & 1 \\
    \hline
    \end{tabular}
\end{center}
\caption{PQ and R symmetry charges of superfields in the example symmetry breaking sector.  The $Q_i$ are in the fundamental representation of a new confining $SU(2)_c$ gauge symmetry, with other fields being singlets under this symmetry.}
\label{table1}
\end{table}
We take a superpotential of the form
\begin{equation}
W \supset \lambda_{ij}^{kl} Z_{kl}Q_iQ_j, 
\end{equation}
With $\lambda_{ij}^{kl}$ anti-symmetric in both upper and lower indices, and zero whenever required by PQ symmetry.
If the $\lambda_{ij}^{kl}$ were zero, then after $SU(2)_c$ confinement at a scale $\Lambda$, one obtains a quantum-corrected modulii space of vacua
parametrized by meson fields  $V_{ij} \sim Q_iQ_j/\Lambda$ satisfying ${\rm Pf}(V) = \Lambda^2$, where  ${\rm Pf}(V)$ is the Pfaffian of $V$ \cite{Seiberg}.
For sufficiently small $\lambda_{ij}^{kl}$,  the condition ${\rm Pf}(V) = \Lambda^2$ is maintained at leading order.  At energies below the
confinement scale one obtains an effective superpotential of the form
\begin{equation}
W \supset \Lambda \lambda_{ij}^{kl} Z_{kl}V_{ij} + X ({\rm Pf}(V)- \Lambda^2), 
\end{equation}
where  X is an auxiliary field introduced to enforce the ${\rm Pf}(V) = \Lambda^2$ constraint.  This constraint, however, cannot be satisfied simultaneously
along with the vanishing of the F-terms of the $Z_{ij}$, and thus supersymmetry is broken.   PQ is also broken for generic
values of the $\lambda_{ij}^{kl}$  by vevs of $V_{12}$ and $V_{34}$.
Interestingly, due to the results of \cite{Witten}, supersymmetry will continue to be broken even if we increase the  $\lambda_{ij}^{kl}$ to become of order 1, and for generic values we would anticipate PQ to also remain broken.
 The $\mu$ term arises due to $W \supset \frac{1}{M_{\rm pl}}Q_1 Q_2 H_U H_D \sim \frac{\Lambda}{M_{\rm pl}} V_{12} H_U H_D$.
Note that a $B \mu$ term is generated of a similar size due to  the supergravity potential containing $-3\frac{\left|W\right|^2}{M_{\rm pl}^2}$, with $W$ having a constant piece $m_{3/2} M_{\rm pl}^2$ to cancel the cosmological constant.
\label{dynamical}

\section{Phenomenology}

Before discussing the testability of our scenario, we must first examine the allowed parameter space for R-parity violation within the theory.  In particular, we will consider constraints which arise from requiring that contributions to neutrino masses be sufficiently small \cite{neutrino}, that baryon asymmetry not be washed out in the early universe \cite{washout}, and that the LSP lifetime be less than about a second to preserve the successful predictions of big bang nucleosynthesis
 (BBN) \cite{BBN}.

In what follows, we will assume bilinear R-parity violation is dominant for simplicity, so that the most important R-parity violating term is $W \supset \mu'_i L_i H_U$, with $i$ a generation index.  This results if we assume that right handed neutrino fields
$N_i$, and therefore also $L_i H_U$ are neutral under $U(1)_{PQ}$, as seems simplest to fit into a grand unified theory framework.   In this context,  other R-parity violating terms may be suppressed by additional powers of the $PQ$ scale.  As
is well known in the literature, one may make a field redefinition of the $L_i$ and $H_D$ so as to absorb the R-parity violating bilinear into the ordinary $\mu$ term.  After doing so, the dominant R-parity violating terms in the superpotential become
\begin{equation}
W \supset \epsilon_i y^d_{jk} L_i Q_j D_k +  \epsilon_i y^e_{jk} L_i L_j E_k,
\end{equation} 
where the $y$'s are the standard Yukawa coupling matrices, and $\epsilon_i = \mu'_i/\mu$.  In what follows, we will discuss the constraints on the $\epsilon_i$, which we will take for simplicity to have a typical value $\epsilon_i \sim \epsilon$.

  The most important upper bounds  come from requiring that the lepton number violating operators do not lead to unacceptably large contributions to neutrino masses, and from requiring that the lepton asymmetry (and through sphaleron processes, also the baryon asymmetry) does not get washed out in the early universe \cite{neutrino, washout}.

The most important contribution to the neutrino masses comes from vacuum expectation values induced in the sneutrinos.  The size of this contribution (which lies in a single direction in neutrino flavor space) is given by
\begin{equation}
\Delta m_\nu = \left(\frac{g'^2}{4 M_1} + \frac{g^2}{4 M_2}\right)\sum_i \left<\tilde{\nu_i}^2\right>,
\end{equation}
with
\begin{equation}
\left<\tilde{\nu}_i\right> = \left(B_i \mu_i' \sin{\beta} +\tilde{m}_{i H_D}^2 \cos{\beta}\right) \frac{v}{m_{\tilde{\nu_i}}^2},
\end{equation}
where $v = 174$GeV, and $B_i \mu_i'$ and $\tilde{m}_{i H_D}^2$ are the coefficients of supersymmetry breaking soft terms $\tilde{L_i} \tilde{H_U}$ and $\tilde{L_i} \tilde{H_D^*}$ respectively, after the $L_i H_U$ superpotential terms have been rotated away.
These soft terms may be expected to be of order $\epsilon m_{3/2}^2$.  It follows that, for roughly universal SUSY scalar masses and gaugino masses of order $\sim m_{\rm gaugino}$,
\begin{equation}
\Delta m_\nu \sim {\rm O}(1) \times 10^{-3} {\rm eV}\left( \frac{\rm TeV}{m_{\rm gaugino}}\right) \left(\frac{\epsilon}{10^{-6}}\right)^2.
\end{equation}
While this expression is rough, and depends on details of various order 1 numerical factors, we take as a rule of thumb that $\epsilon \lesssim 10^{-6}$ is preferred in order to avoid fine tuning in the neutrino mass matrix.

Avoiding washout of the lepton asymmetry requires that the lepton number violating interactions are out of equilibrium in the early universe.  This constraint was worked out in \cite{washout}, and requires roughly
\begin{equation}
\epsilon \lesssim {\rm O}(1)  \times 10^{-5} \left(\frac{3}{\tan{\beta}}\right)\left(\frac{M_{SUSY}}{100 {\rm TeV}}\right)^{1/2}.
\end{equation}

We will next discuss the constraints arising from BBN \cite{BBN}.  We must require that the lifetime of the LSP is less than about a second.  From equations (\ref{eq:gluino})-(\ref{eq:bino}), whether or not the LSP is the bino, wino, or gluino depends on both $L$ and $M_{SUSY}$.  The lifetimes of wino or bino LSPs are given by
\begin{eqnarray}
\tau_{\rm wino} &\simeq& \frac{16 \pi}{g_2^2 \epsilon^2 m_{\rm wino} } \sim \left(\frac{10^{-12}}{\epsilon}\right)^2 \frac{ {\rm TeV}}{m_{\rm wino}}\times 10^{-1}{\rm  sec},\\ 
\tau_{\rm bino} &\simeq&   \frac{16 \pi}{g_1^2 \epsilon^2 m_{\rm bino} } \sim \left(\frac{10^{-12}}{\epsilon}\right)^2 \frac{ {\rm TeV}}{m_{\rm bino}}\times 10^{-1}{\rm  sec} .
\end{eqnarray}
As a rule of thumb, we should have $\epsilon \gtrsim 10^{-12}$ in order to satisfy BBN constraints for a wino or bino LSP.

For a gluino LSP, the lifetime is given by
\begin{equation}
\tau_{\rm gluino} \simeq \frac{128 \pi^2}{\alpha_{QCD} \epsilon^2 \lambda_b^2} \frac{M_{SUSY}^4}{m_{\rm gluino}^5} \sim  {\rm O}(1)\ \times .1\ {\rm sec} \times \left(\frac{10^{-6}}{\epsilon}\right)^2 \left(\frac{3}{\tan{\beta}}\right)^2\left(\frac{M_{SUSY}}{100 {\rm TeV}}\right)^4\left(\frac{1 {\rm TeV}}{m_{\rm gluino}}\right)^5 .
\label{gluinolife}
\end{equation}
In this case however, the final relic abundance has great uncertainty \cite{gluinorelic1, gluinorelic2}.  This is due to the fact that after the QCD phase transition, the gluinos form QCD bound states, with strong interaction processes now playing a role in their annihilation.
While a gluino LSP naively requires  $\epsilon \gtrsim 10^{-6}$, it is interesting to note that, for a sufficiently suppressed relic abundance and $\epsilon$ of perhaps $10^{-7}$, gluino decays could conceivably play a role in addressing the cosmological lithium abundance problems \cite{lithiumobs, lithium}.  We note, however, that a gluino LSP requires a somewhat large value of $L$, with $L/m_{3/2} \gtrsim 20$.

For the most likely cases of a wino or bino LSP, our upper and lower bounds on the $\epsilon_i$ thus require that these parameters be between roughly $10^{-12}$ and $10^{-6}$, while for a gluino LSP, $\epsilon_i$ of order $10^{-6}$ is needed.  The precise size of the allowed window depends on the details of the superpartner spectrum, and also widens as $\Lambda_{SUSY}$ is increased.  If $\Lambda_{SUSY}$ is $\sim 10^{13}$ GeV, for example, then the range for the $\epsilon_i$ in the case of a wino or bino LSP becomes perhaps $10^{-13}-10^{-5}$.   We thus conclude that
our scenario is viable for a rather broad range in the amount of R-parity violation.

We will now move on to address the testability of our model.  The basic predictions are axionic dark matter, gauginos with order $1-10$ TeV masses, a PQ scale of the order of the SUSY breaking scale, and R-parity violation.

Axionic dark matter, with $f_a$ in roughly the range $6\times 10^{11} - 6\times 10^{12}$ GeV and DFSZ-like PQ charges \cite{DFSZ} as in our model, is expected to be accessible by phase II of the ADMX microwave cavity search experiment \cite{ADMX}.  It follows that we  would anticipate seeing a signal there unless the initial displacement angle $\Theta_i$ is smaller than about $\pi/10$.  This corresponds to the ability to probe PQ/SUSY breaking scales up to a little over $10^{13}$ GeV, 
or gravitino masses of perhaps tens of thousands of TeV.

The collider phenomenology in our model depends on the amount of R-parity violation present.  For wino or bino LSPs, the 
decay lengths are 
\begin{eqnarray}
c \tau_{\rm wino} &\simeq&\left(\frac{10^{-6}}{\epsilon}\right)^2 \frac{ {\rm TeV}}{m_{\rm wino}}\times 10^{-3}{\rm  cm},\\ 
c \tau_{\rm bino} &\simeq&    \left(\frac{10^{-6}}{\epsilon}\right)^2 \frac{ {\rm TeV}}{m_{\rm bino}}\times 10^{-2}{\rm  cm} .
\end{eqnarray}
Thus for $\epsilon$ near its upper bound, wino or bino LSPs would decay quickly inside the detector.  For small $\epsilon$
these LSPs become stable on collider length scales, while for intermediate values of $\epsilon$ there is a possibility for
displaced vertex signatures.  
Although perhaps disfavored by the large values of $L$ required ($L/m_{3/2} \gtrsim 20$), if the LSP is the gluino, the
decay length is necessarily long (c.f. equation (\ref{gluinolife})).  On the other hand, after production, the gluino hadronizes and the signatures depend on the spectroscopy of the resulting hadrons (most importantly on the ratio of charged to neutral gluino-hadrons).  We will not review all of the relevant phenomenology here, but signatures may include monojets, detection of anomalously heavy charged particles, or stopped late decaying gluinos \cite{gluinocollide}.  The range accessible to the LHC may be up to about 2.4 TeV, with  current limits excluding masses up to about 1 TeV.  Finding evidence of a gluino LSP would generally point towards R-parity violation, since a gluino LSP would have contributed strongly interacting dark matter particles in conflict with experiment.  This evidence would of course become more explicit  if one could directly observe decays of stopped gluinos.\footnote{Long lived non-LSP gluinos, as in the split SUSY scenario with very heavy scalars could hopefully be distinguished by a lack of events involving direct LSP wino production signals.  In our case with a gluino LSP, such directly produced winos would quickly decay through standard MSSM interactions.}

For wino LSPs with long lifetimes, the collider phenomenology is similar to that of the standard R-parity conserving pure gravity mediation model, as discussed in  \cite{LHCPGM}.   There it was found that if the gluino is light enough to be sufficiently produced at the LHC, then it should be possible to see a signal in jets + missing energy.  For the cases in which a charged wino is produced in the final state, it should be possible to see in addition a disappearing charged track signal.  The former search, ignoring the charged tracks,  and conducted at the 14 TeV LHC, is projected to be sensitive to gluino masses less than $\sim 2.4$ TeV and wino masses less than $\sim 1$ TeV.  The current limit has $m_{\rm gluino} < 1$ TeV and ${m_{\rm wino} < 300}$ GeV.   If the gluino is beyond the reach of the LHC, then it may still be possible to find a signal of a long lived wino LSP  through direct production, and looking for disappearing charged track events, although the precise sensitivity cannot be determined at this time due to the difficulty of estimating backgrounds.   
  An interesting possible signature of our model could arise if a long lived wino LSP were discovered, but without a corresponding signature in cosmic ray searches for dark matter.  Indeed, the strictest present constraints on a stable wino LSP of $m_{\rm wino} \gtrsim 300$ GeV come from such indirect dark matter detection results \cite{Ibe2},\footnote{For a more recent analysis obtaining a more stringent constraint of perhaps  $m_{\rm wino} \gtrsim 500$ GeV, see \cite{Hooper}.  This result depends on some 
background subtractions from the galactic center, and is perhaps not as completely robust as the result of \cite{Ibe2}.} and these would be avoided in our scenario due to the axion forming the dark matter.  As such, wino masses down to about 100 GeV are still allowed in our scenario.

If the bino is the LSP and is long lived, then the same jets + missing energy searches as in the wino case will be applicable assuming that the gluino is within reach.  Interestingly, it should be possible to distinguish the bino LSP case from the wino case by noting an $\it absence$ of disappearing charged track events.  This situation would then suggest that R-parity is violated, since a completely stable bino LSP would have yielded too much dark matter in the early universe.

For short lived wino or bino LSPs the collider signatures become rich, including events with multiple isolated charged leptons,
as well as possible displaced vertices.  These striking signatures would make the model relatively easy to look for due to greatly reduced standard model backgrounds.  This leaves open an exciting possibility:  It may be possible to test our 
model very explicitly, by observing direct signatures of both axionic dark matter, as well as the presence of R-parity violating
decays of gauginos with masses near the weak scale!  We leave a complete study of the collider phenomenology of our 
model in the case of short lived wino or bino LSPs for future work, but please see for example 
\cite{RPVpapers} for prior investigations of 
 R-parity violating neutralino decays.

\label{pheno}

\section{Summary}
In this paper we have considered a simple scenario with a somewhat high supersymmetry parter scale $\gtrsim 100$ TeV, and attempted to explain why the weak scale might be somewhat natural, but not completely so.

We have taken the strong CP problem as a starting point, and added an axion to the theory.  In addition, we have hypothesized that the PQ and SUSY breaking scales might be connected, and therefore of a similar size.  Together with an assumption of R-parity violation, the axion is required to compose dark matter and seed structure formation.  This leads to a constraint  $\Lambda_{SUSY} \sim \Lambda_{PQ} \gtrsim 10^{12}$ GeV, with a superpartner mass scale of $\gtrsim 100$ TeV.  The theory will be tested at upcoming experiments:  In particular, we would expect to see a positive signal at the ADMX dark matter axion search experiment.  In addition, the gauginos have a good chance to be accessible at the LHC, due to their loop-factor suppressed masses.  A gluino or bino LSP would signal that R-parity is violated, in order to avoid cosmological constraints.  A wino LSP, if sufficiently light, could also point towards R-parity violation, since we predict an absence of  a dark matter-annihilation cosmic ray signature.   It may even be possible to directly observe R-parity violating decays of the LSP, thereby explicitly confirming our scenario.

We gave one example of a dynamical  symmetry breaking sector which can relate the PQ and SUSY scales.  In this model, there is actually a free parameter which sets the ratio of $\Lambda_{PQ}$ to $\Lambda_{SUSY}$, and which we took to be of order 1 by assumption.  It would be interesting to try to construct a 
model with no such free parameters, which could set the ratio of these two scales in a completely dynamical way.  We will leave this question open for future work.

Finally, there is perhaps a  general aspect of our scenario which deserves further emphasis:  The presently available information about the weak scale- an apparently elementary Higgs boson, no signals yet of new physics, and the success of supersymmetric gauge coupling unification- suggests that supersymmetry may be realized in nature in a  way which requires significant fine tuning, but yet which does address the bulk of the hierarchy problem.  This presents a puzzle:  Why might nature prefer an \emph{almost} natural- but not \emph{completely} natural- weak scale?  In this paper we have shown that this puzzle might be of practical importance in suggesting directions for phenomenology.  Our attempted explanation makes concrete predictions, and it is
conceivable that there might be other solutions with their own. 
\label{conclusions}

\section*{Acknowledgments}

This work is supported by the Grant-in-Aid for Scientific research from the Ministry of Education, Science, Sports, and Culture (MEXT), Japan, No. 24740151 (M.I.), No. 22244021 (T.T.Y.) and also by the World Premier International Research Center Initiative (WPI Initiative), MEXT, Japan.



\begin{thebibliography}{99}

\bibitem{anomaly} 
  G.~F.~Giudice, M.~A.~Luty, H.~Murayama and R.~Rattazzi,
  JHEP {\bf 9812}, 027 (1998)
  [hep-ph/9810442].

\bibitem{Ibe1} 
M.~Ibe and T.~T.~Yanagida,
Phys.\ Lett.\ B {\bf 709}, 374 (2012).

\bibitem{Ibe2}
M.~Ibe, S.~Matsumoto and T.~T.~Yanagida,
Phys.\ Rev.\ D {\bf 85}, 095011 (2012).

\bibitem{LHCPGM} 
  B.~Bhattacherjee, B.~Feldstein, M.~Ibe, S.~Matsumoto and T.~T.~Yanagida,
  arXiv:1207.5453 [hep-ph].

\bibitem{PQ} 
  R.~D.~Peccei and H.~R.~Quinn,
  Phys.\ Rev.\ Lett.\  {\bf 38}, 1440 (1977);
  R.~D.~Peccei and H.~R.~Quinn,
  Phys.\ Rev.\ D {\bf 16}, 1791 (1977);
  S.~Weinberg,
  Phys.\ Rev.\ Lett.\  {\bf 40}, 223 (1978);
  F.~Wilczek,
  Phys.\ Rev.\ Lett.\  {\bf 40}, 279 (1978).

\bibitem{PQmu} 
  J.~E.~Kim and H.~P.~Nilles,
  Phys.\ Lett.\ B {\bf 138}, 150 (1984);
  E.~J.~Chun, J.~E.~Kim and H.~P.~Nilles,
  Nucl.\ Phys.\ B {\bf 370}, 105 (1992);
  B.~Feldstein, L.~J.~Hall and T.~Watari,
  Phys.\ Lett.\ B {\bf 607}, 155 (2005)
  [hep-ph/0411013].

\bibitem{PQDM} 
  L.~F.~Abbott and P.~Sikivie,
  Phys.\ Lett.\ B {\bf 120}, 133 (1983);
  J.~Preskill, M.~B.~Wise and F.~Wilczek,
  Phys.\ Lett.\ B {\bf 120}, 127 (1983);
  M.~Dine and W.~Fischler,
  Phys.\ Lett.\ B {\bf 120}, 137 (1983);
  P.~Sikivie,
  Lect.\ Notes Phys.\  {\bf 741}, 19 (2008)
  [astro-ph/0610440].
  L.~D.~Duffy and K.~van Bibber,
  New J.\ Phys.\  {\bf 11}, 105008 (2009)
  [arXiv:0904.3346 [hep-ph]].

\bibitem{DMreq}
  S.~Hellerman and J.~Walcher,
  Phys.\ Rev.\ D {\bf 72}, 123520 (2005)
  [hep-th/0508161];
  M.~Tegmark, A.~Aguirre, M.~Rees and F.~Wilczek,
  Phys.\ Rev.\ D {\bf 73}, 023505 (2006)
  [astro-ph/0511774].

\bibitem{walls} 
  P.~Sikivie,
  Phys.\ Rev.\ Lett.\  {\bf 48}, 1156 (1982).

\bibitem{dynamical} 
  K.~-I.~Izawa and T.~Yanagida,
  Prog.\ Theor.\ Phys.\  {\bf 95}, 829 (1996)
  [hep-th/9602180];
  K.~A.~Intriligator and S.~D.~Thomas,
  Nucl.\ Phys.\ B {\bf 473}, 121 (1996)
  [hep-th/9603158].


\bibitem{Seiberg} 
  N.~Seiberg,
  Phys.\ Rev.\ D {\bf 49}, 6857 (1994)
  [hep-th/9402044];
  N.~Seiberg,
  Phys.\ Lett.\ B {\bf 318}, 469 (1993)
  [hep-ph/9309335].

\bibitem{Witten} 
  E.~Witten,
  Nucl.\ Phys.\ B {\bf 202}, 253 (1982).

\bibitem{neutrino} 
  D.~E.~Kaplan and A.~E.~Nelson,
  JHEP {\bf 0001}, 033 (2000)
  [hep-ph/9901254].

\bibitem{washout} 
  B.~A.~Campbell, S.~Davidson, J.~R.~Ellis and K.~A.~Olive,
  Phys.\ Lett.\ B {\bf 256}, 457 (1991);
  W.~Fischler, G.~F.~Giudice, R.~G.~Leigh and S.~Paban,
  Phys.\ Lett.\ B {\bf 258}, 45 (1991);
  H.~K.~Dreiner and G.~G.~Ross,
  Nucl.\ Phys.\ B {\bf 410}, 188 (1993)
  [hep-ph/9207221].

\bibitem{BBN} 
  M.~Kawasaki, K.~Kohri and T.~Moroi,
  Phys.\ Lett.\ B {\bf 625}, 7 (2005)
  [astro-ph/0402490].
  M.~Kawasaki, K.~Kohri and T.~Moroi,
  Phys.\ Rev.\ D {\bf 71}, 083502 (2005)
  [astro-ph/0408426].

\bibitem{gluinorelic1} 
  H.~Baer, K.~-m.~Cheung and J.~F.~Gunion,
  Phys.\ Rev.\ D {\bf 59}, 075002 (1999)
  [hep-ph/9806361].

\bibitem{gluinorelic2} 
  A.~Arvanitaki, C.~Davis, P.~W.~Graham, A.~Pierce and J.~G.~Wacker,
  Phys.\ Rev.\ D {\bf 72}, 075011 (2005)
  [hep-ph/0504210].

\bibitem{lithiumobs} 
  P.~Bonifacio, P.~Molaro, T.~Sivarani, R.~Cayrel, M.~Spite, F.~Spite, B.~Plez and J.~Andersen {\it et al.},
  [astro-ph/0610245];
  S.~G.~Ryan, T.~C.~Beers, K.~A.~Olive, B.~D.~Fields and J.~E.~Norris,
  Astrophys.\ J.\  {\bf 530}, L57 (2000);
  J.~Melendez and I.~Ramirez,
  Astrophys.\ J.\  {\bf 615}, L33 (2004)
  [astro-ph/0409383];
  M.~Asplund, D.~L.~Lambert, P.~E.~Nissen, F.~Primas and V.~V.~Smith,
  Astrophys.\ J.\  {\bf 644}, 229 (2006)
  [astro-ph/0510636].

\bibitem{lithium} 
  K.~Jedamzik,
  Phys.\ Rev.\ D {\bf 70}, 063524 (2004)
  [astro-ph/0402344];
  K.~Kohri, T.~Moroi and A.~Yotsuyanagi,
  Phys.\ Rev.\ D {\bf 73}, 123511 (2006)
  [hep-ph/0507245];
\bibitem{Cumberbatch:2007me} 
  D.~Cumberbatch, K.~Ichikawa, M.~Kawasaki, K.~Kohri, J.~Silk and G.~D.~Starkman,
  Phys.\ Rev.\ D {\bf 76}, 123005 (2007)
  [arXiv:0708.0095 [astro-ph]].


\bibitem{wallsolve} 
  H.~Georgi and M.~B.~Wise,
  Phys.\ Lett.\ B {\bf 116}, 123 (1982);
  L.~F.~Abbott and P.~Sikivie,
  Phys.\ Lett.\ B {\bf 120}, 133 (1983).

\bibitem{DFSZ} 
  M.~Dine, W.~Fischler and M.~Srednicki,
  Phys.\ Lett.\ B {\bf 104}, 199 (1981);
  A.~R.~Zhitnitsky,
  Sov.\ J.\ Nucl.\ Phys.\  {\bf 31}, 260 (1980)
  [Yad.\ Fiz.\  {\bf 31}, 497 (1980)].

\bibitem{ADMX} 
  S.~J.~Asztalos {\it et al.}  [ADMX Collaboration],
  Phys.\ Rev.\ Lett.\  {\bf 104}, 041301 (2010)
  [arXiv:0910.5914 [astro-ph.CO]];
  S.~J.~Asztalos, R.~Bradley, G.~Carosi, J.~Clarke, C.~Hagmann, J.~Hoskins, M.~Hotz and D.~Kinion {\it et al.},

\bibitem{Hooper} 
  D.~Hooper, C.~Kelso and F.~S.~Queiroz,
  arXiv:1209.3015 [astro-ph.HE].

\bibitem{gluinocollide} 
  J.~L.~Hewett, B.~Lillie, M.~Masip and T.~G.~Rizzo,
  JHEP {\bf 0409}, 070 (2004)
  [hep-ph/0408248];
  A.~Arvanitaki, S.~Dimopoulos, A.~Pierce, S.~Rajendran and J.~G.~Wacker,
  Phys.\ Rev.\ D {\bf 76}, 055007 (2007)
  [hep-ph/0506242].

\bibitem{RPVpapers}
  F.~de Campos, O.~J.~P.~Eboli, M.~B.~Magro and D.~Restrepo,
  Phys.\ Rev.\ D {\bf 79}, 055008 (2009)
  [arXiv:0809.0007 [hep-ph]].

  G.~Aad {\it et al.}  [ATLAS Collaboration],
  Phys.\ Lett.\ B {\bf 707}, 478 (2012)
  [arXiv:1109.2242 [hep-ex]].

  G.~Aad {\it et al.}  [ATLAS Collaboration],
  Phys.\ Rev.\ D {\bf 85}, 012006 (2012)
  [arXiv:1109.6606 [hep-ex]].

  P.~W.~Graham, D.~E.~Kaplan, S.~Rajendran and P.~Saraswat,
  JHEP {\bf 1207}, 149 (2012)
  [arXiv:1204.6038 [hep-ph]].

  G.~Aad {\it et al.}  [ATLAS Collaboration],
  arXiv:1210.7451 [hep-ex].



\end{thebibliography}
\end{document}